\documentclass[3p,review]{elsarticle}

\usepackage{lineno,hyperref}
\usepackage{natbib}
\usepackage{adjustbox}
\usepackage{longtable}
\usepackage{booktabs}
\usepackage{xcolor}
\usepackage{caption}
\usepackage{rotating}
\usepackage{lscape}
\usepackage{soul}
\usepackage{textcomp}
\usepackage{multirow}
\usepackage{multicol}
\usepackage{tabu}
\usepackage{tocloft}
\usepackage{blindtext}
\usepackage{mathtools}
\usepackage{blkarray}
\usepackage{subcaption}
\usepackage{float}
\usepackage{array}
\usepackage{amsmath}
\usepackage{amssymb}
\usepackage{setspace}
\usepackage{bm}
\usepackage{algorithm}
\usepackage{algpseudocode}
\usepackage{makecell}
\usepackage{siunitx}
\DeclareUnicodeCharacter{2212}{-}

\begin{document}

\begin{frontmatter}

\title{Multi-State Modeling of Greenhouse Cucumber Yield Dynamics Under Microclimate Effects}


\author[1]{\corref{cor1} Emiliano Seri}
\ead{emiliano.seri@uniroma2.it}
\cortext[cor1]{Corresponding author}

\author[1]{Francesco Biso}
\ead{francesco.biso@uniroma2.it}

\author[1]{Gianluigi Bovesecchi}
\ead{gianluigi.bovesecchi@uniroma2.it}

\author[2]{Nikolaos Katsoulas}
\ead{nkatsoul@uth.gr}

\author[1]{Cristina Cornaro}
\ead{cristina.cornaro@uniroma2.it}

\address[1]{Department of Enterprise Engineering, University of Rome Tor Vergata, Via del Politecnico 1, 00133, Rome, Italy}

\address[2]{Lab of Agricultural Constructions and Environmental Control, Department of Agriculture Crop Production and Rural Environment, University of Thessaly, Fytokou Str, 38446, Volos, Greece}

\begin{abstract}
Greenhouse decisions often rely on static thresholds, yet crop output switches among production regimes that respond nonlinearly to microclimate. We model daily cucumber production as transitions among three ordered yield states (low, medium, high) using a continuous-time,
covariate-dependent multistate (Markov) model to quantify how fast crop lines migrate between states and how microclimate and structural features accelerate or decelerate those moves. Data come from four adjacent greenhouse compartments in Volos, Greece (six lines per greenhouse compartment; 62 days, Oct--Dec 2024). States are defined once from the control greenhouse (GH-3) tertiles and applied unchanged across houses to preserve agronomic comparability. Transition intensities depend on within-greenhouse $z$-scores of relative humidity (RH), photosynthetically active radiation (PAR) and CO\textsubscript{2}, plus greenhouse fixed effects.

Results reveal a coherent dynamic: an inherent upward drift through the medium state, “sticky’’ low-yield spells unless conditions improve, and short-horizon persistence once high yield is reached. Across houses, RH and PAR are the dominant, interpretable levers. Humid or brighter days systematically accelerate upgrades and suppress regressions, whereas day-to-day CO\textsubscript{2} deviations show no clear pooled signal once RH and PAR are included. Residual differences between greenhouses are modest and statistically uncertain over this horizon.

By mapping intensities to short-term transition probabilities and expected sojourns, the model turns sensor streams into actionable guidance: when raising RH most effectively exits low/medium, where to concentrate supplemental light to push medium to high, and how likely a high state is to persist without intervention. The approach provides an interpretable bridge between microclimate control and near-term production and a lightweight building block for greenhouse digital twins enabling line ranking by upgrade risk, valuation of feasible set-point changes, and scenario analysis under alternative humidity and lighting strategies.
\end{abstract}

\begin{highlights}
\item Cucumber yield modelled as transitions among three ordered states.
\item RH and PAR strongly speed upgrades and slow regressions.
\item CO$_2$ day-to-day deviations show no clear pooled effect.
\item Intensities yield 7--30 day upgrade probabilities and waiting times.
\item Cross-house pooling shows modest baselines; focus on daily steering.
\end{highlights}


\end{frontmatter}


\section{Introduction}
\label{sec:intro}

Modern greenhouses are data-rich, sensor-dense environments:
CO\textsubscript{2}, air temperature, relative humidity (RH) and photosynthetically active radiation (PAR) are logged every few minutes; fertigation volumes are tallied daily; on–line weighing systems record fruit growth almost in real time.  
Yet, despite the stochastic and interacting nature of these data
streams, horticultural decision support still relies largely on static thresholds or simple linear regressions.  
A statistical framework able to respect irregular observation times, abrupt growth spurts and covariate-dependent dynamics is urgently needed.

A wide spectrum of empirical and statistical models has been used to predict crop yields.  
Comparative studies assess parametric response functions, generalized additive models (GAM), multi-linear regressions, and machine-learning methods such as Random Forests \citep{hegedus2023assessing}.  
Time-series approaches are also common, from classical ARIMA-type tools \citep{michel2013comparison,rathod2021two} to autoregressive distributed-lag (ARDL) models for quantifying climate impacts on agricultural output \citep{abebe2024influence}.  
Bayesian frameworks are frequently adopted to analyze yield responses \citep{cho2023bayesian}, including MCMC-based studies of drought effects on wheat \citep{yildirak2015bayesian}.  
Within greenhouses, cucumber yield has been related to microclimate using multiple linear regression, typically highlighting total radiation and air temperature as key drivers \citep{Ding_2019_hortsci}, and sigmoidal growth curves have been fitted for other protected crops \citep{horticulturae8111021}.  
However, these models usually target a continuous outcome or a single end point and do not directly characterize transitions between distinct production regimes.

We adopt continuous-time \textit{multi-state modelling} (MSM) as a complementary lens.  
MSMs are well established in biology and ecology, including plant demography where stage-structured transitions are modelled explicitly \citep{GREGG200650,kery2004demographic}, but their use for greenhouse yield dynamics is still limited and, to the best of our knowledge, no prior greenhouse yield study has used this framework.
An early study that employed a Markov chain approach to forecast crop yields was provided by \cite{matis1985markov}. They categorized crop yield outcomes into discrete states and used transition probabilities to predict yield distributions during the growing season. This was applied in a field context, using soil moisture data to drive transitions. They achieved reasonable pre-harvest yield forecasts by simulating state transitions, highlighting the potential of Markov models for crop yield prediction.
The follow-up work \cite{matis1989application}, extended the Markov chain approach for yields, and demonstrated that discrete-state models can capture yield dynamics and uncertainties over time.
Here we illustrate MSMs on a hydroponic cucumber (\emph{Cucumis sativus}) experiment conducted in four adjacent
greenhouse compartments (GH-2–GH-5) at the University of Thessaly, Greece.
Daily fruit weight is discretised into three ordered yield states.  The greenhouses by treatment: GH-3 is the control (reference microclimate), GH-2 applies CO\textsubscript{2} enrichment, GH-4 has a PV roof, and GH-5 combines a PV roof with CO\textsubscript{2} enrichment.
The choice of three states is a pragmatic balance: tertiles derived from the control house (GH-3) delimit (i) a low-yield phase dominated by vegetative growth, (ii) a mid-yield plateau where most production occurs, and (iii) an upper tail of short-lived but agronomically meaningful peaks.  
Finer splits would inflate sampling error, whereas a binary split would mask the biologically distinct “take-off” and “peak” phases.

Our primary goal is to \emph{quantify how quickly crop lines migrate between these yield states and how CO\textsubscript{2}, RH and PAR accelerate or decelerate the moves}.  
Hazard ratios extracted from the MSM translate sensor readings into actionable odds (e.g., “a +1 SD PAR day makes the jump from mid to high yield six times faster”).

The analysis proceeds in two stages.  
First, an MSM is fitted to the control house GH-3 to develop the
workflow and expose data idiosyncrasies.  
Second, all four houses are pooled into a single model with fixed
greenhouse effects, boosting statistical power and enabling formal
cross-house comparison.  
This study, conducted within the framework of the European
REGACE project, is a step toward developing a comprehensive \emph{digital twin} for PV-integrated greenhouses, complementing the microclimate prediction module already deployed in \citet{seri2025greenhouse}.

Results reveal a pronounced upward drift: under baseline conditions a mid-yield plot needs about 23 days to reach the high-yield state, but a one-standard-deviation PAR increase shortens this to fewer than five days, while elevated RH is the main lever for escaping low yield.
Across houses, fixed effects differ by less than two-fold, suggesting that the MSM captures shared physiological behaviour with modest structural adjustment.

The remainder of the paper is organised as follows.
Section~\ref{sec: Data} describes the experimental setup, data sources and preprocessing;
Section~\ref{sec: method} details the MSM formulation and estimation strategy;
Section~\ref{sec: Analysis} reports results for the single-house and pooled models, supplemented by diagnostic visualisations;
Section~\ref{sec: Discussion} discusses agronomic implications,
limitations and avenues for hierarchical or real-time extensions;
finally, Section~\ref{sec: conclusions} summarises the key findings.

\section{Data}
\label{sec: Data}

The trial was carried out in four adjacent greenhouse compartments located in Volos, Greece.  
All greenhouses share the same geometry, hydroponic gutters and climate controller, but differ in the single factor under study:

\begin{center}
\begin{tabular}{@{}lcl@{}}
\toprule
House & Code & Treatment \\ \midrule
Greenhouse 3 & GH-3 & Control (reference microclimate)\\
Greenhouse 2 & GH-2 & \(\mathrm{CO_2}\) enrichment\\
Greenhouse 4 & GH-4 & PV roof\\
Greenhouse 5 & GH-5 & PV \(+\)\; \(\mathrm{CO_2}\) enrichment\\
\bottomrule
\end{tabular}
\end{center}

Each greenhouse hosts six crop lines (\textit{Line 1–6}), giving \(4\times6=24\) experimental units followed for 62 days (10~Oct–10 Dec 2024).

\subsection{Data streams and pre-processing}

Three complementary data streams were collected.  
First, the climate controller logged air temperature, relative humidity (RH), photosynthetically active radiation (PAR) and \(\mathrm{CO_2}\) concentration every 15 min.  
Second, weighing tanks provided daily irrigation and drainage volumes, from which net water balance ($L\,m^{-2}$) was derived.  
Third, weekly cucumber harvest weights were recorded for each of the six crop lines in every house.

All climate variables were time-ordered and averaged from 15-min values to hourly and then to daily means.  
Occasional gaps in the 15-min log (\(<6\%\) of entries) were filled with a \emph{three-day neighbourhood} smoother: a missing value inherits the mean of the same time slot at \(\pm1,\pm2,\pm3\) days, thereby honouring the diurnal cycle while borrowing strength from adjacent days.  
The water-balance series contained no missing values, and weekly harvest weights were available for every crop line and week; these complete values were simply forward-carried to produce a daily yield vector for modelling.

After these steps the analysis table contains 24 subjects (4 houses \(\times\) 6 lines) observed on 62 consecutive days, giving \(24\times62 = 1{,}488\) daily records with the variables summarised in Table \ref{tab:variables}.

\begin{table}[H]
\centering\small
\caption{Daily variables entering the multi–state analysis}
\label{tab:variables}
\begin{tabular}{@{}llp{7cm}@{}}
\toprule
Symbol / name      & Units & Description \\ \midrule
$\text{DayIndex}$  & days & Study day counted from 10 Oct to 10 Dec 2024. \\
$Y_{it}$           & \si{kg\,plant^{-1}} & Fresh cucumber weight of crop line $i$ on day $t$. \\[2pt]
$\text{CO\textsubscript{2}}$ & \si{ppm} & Mean greenhouse \(\mathrm{CO_2}\) concentration (15-min log $\rightarrow$ daily mean). \\[2pt]
$\mathrm{RH}$      & \%  & Mean relative humidity. \\
$\mathrm{PAR}$     & \si{\micro mol\,m^{-2}\,s^{-1}} & Mean photosynthetically active radiation. \\[2pt]
$\mathrm{Temp}$    & \si{\degreeCelsius} & Mean air temperature \textit{(discarded due to collinearity with PAR)}. \\[2pt]
$\mathrm{Water}$   & \si{L\,m^{-2}} & Net irrigation $-$ drainage \textit{(discarded; recirculating hydroponics)}. \\[2pt]
$X_{it}$           & --  & Discrete yield state (1$=$low, 2$=$medium, 3$=$high). \\ 
\bottomrule
\end{tabular}
\end{table}

\subsection{Variable screening and normalization}

\begin{figure}[H]
  \centering
  \caption{Daily Pearson correlations of detrended within-greenhouse anomalies (residuals from a smooth day-of-season trend); blue = positive, red = negative.}
  \begin{subfigure}{0.495\linewidth}
  \caption{GH-3 (control)}\label{fig:corrGH3}
    \includegraphics[width=\linewidth]{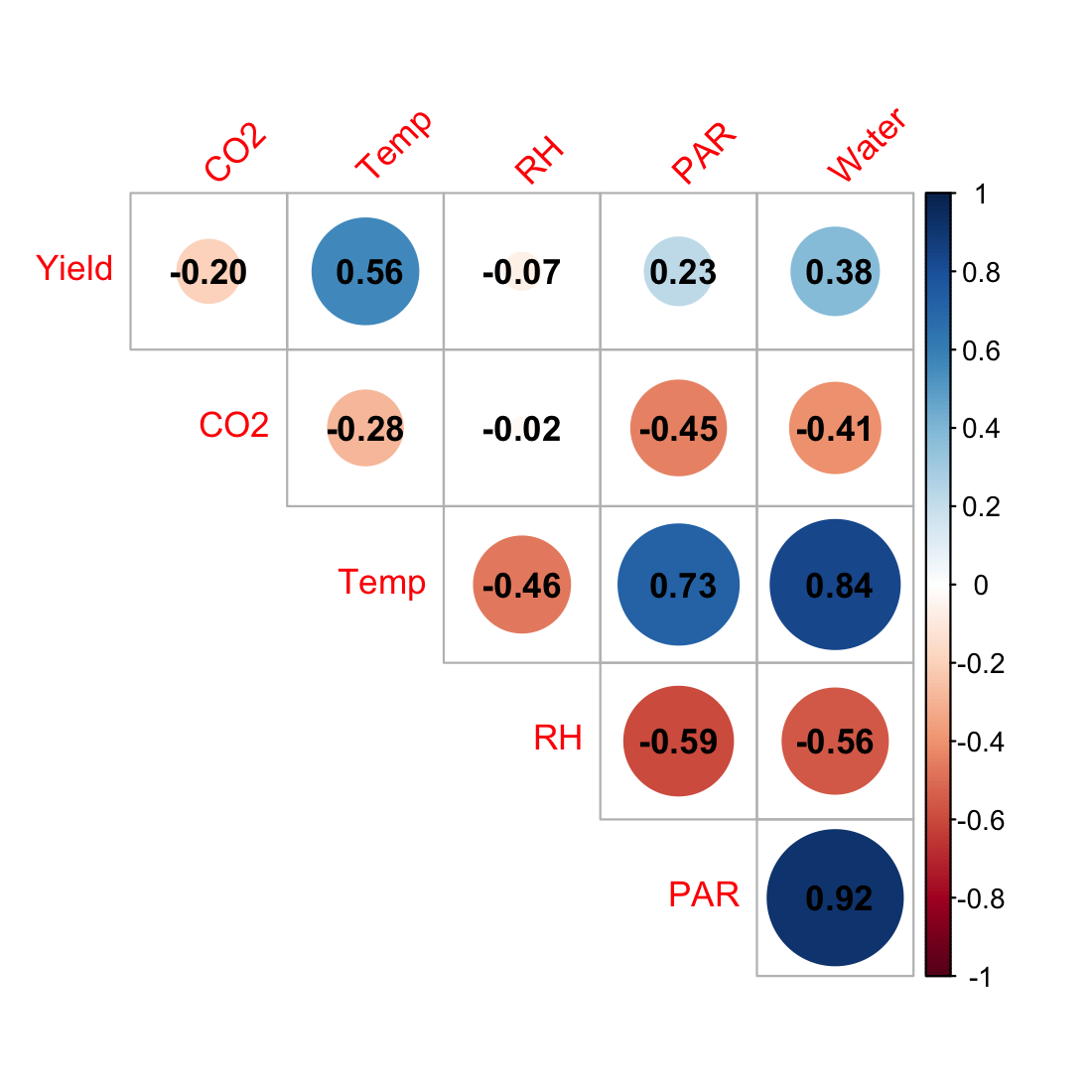}
  \end{subfigure}\hfill
  \begin{subfigure}{0.495\linewidth}
  \caption{All greenhouses pooled.}\label{fig:corrAll}
    \includegraphics[width=\linewidth]{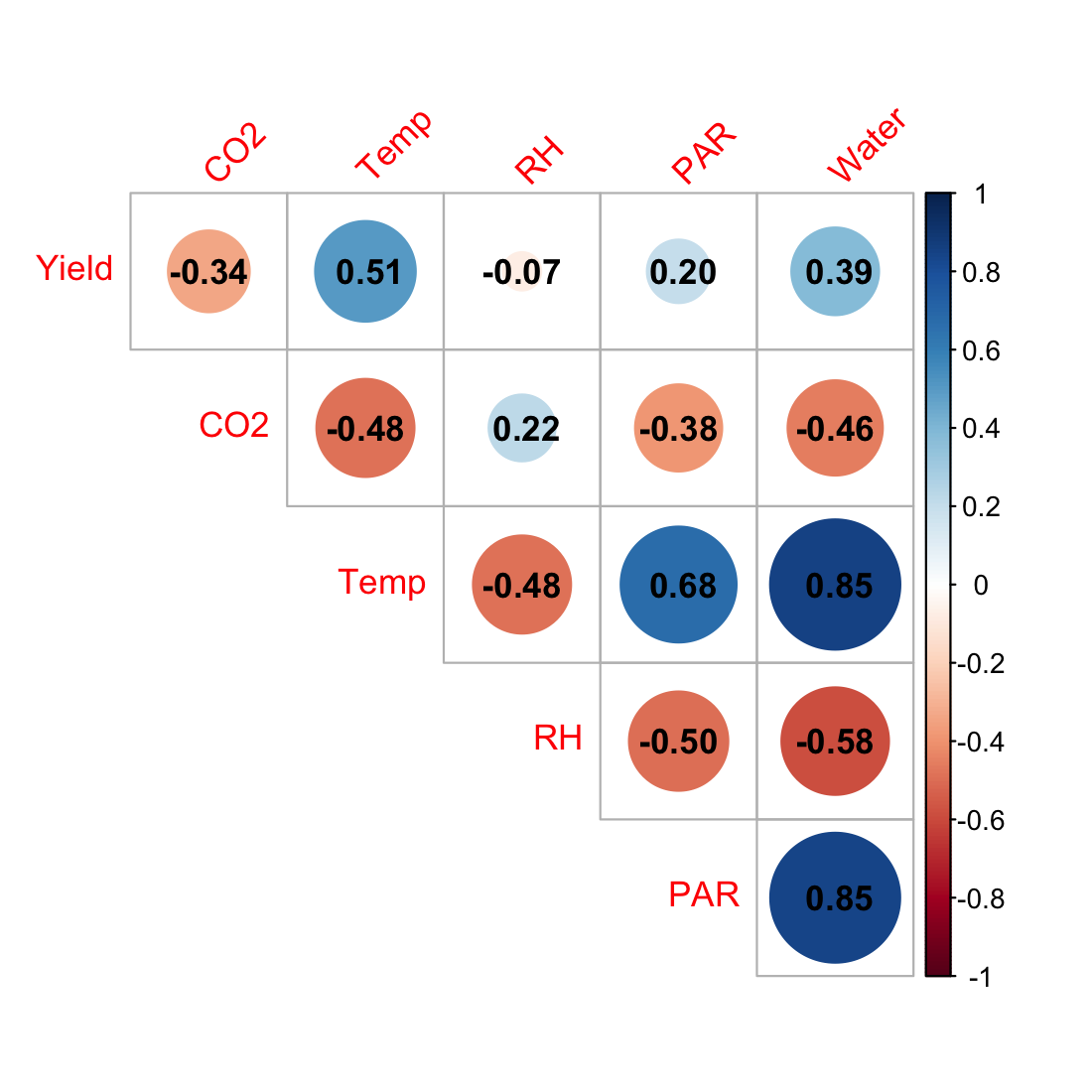}
  \end{subfigure}
\end{figure}

Figures \ref{fig:corrGH3} (GH-3 only) and \ref{fig:corrAll} (all greenhouses combined) report pairwise Pearson correlations computed on detrended, within-greenhouse anomalies (residuals from a smooth day-of-season function), to avoid spurious associations driven by common trends/seasonality in daily climate series. After detrending, Temperature and net Water remain strongly correlated with PAR (as expected: radiation drives heating and evapotranspiration). Given the limited number of experimental units (six lines in GH-3; 24 lines overall) and the sparse number of events per transition in the MSM, retaining all highly correlated covariates would inflate standard errors and jeopardize identifiability (near-singular Hessians, unstable estimates). We therefore adopt parsimonious, physiologically distinct set and exclude Temperature and Water, retaining CO\textsubscript{2}, RH and PAR.

Note that the experiment uses \emph{recirculating hydroponics}; the net
Water balance primarily reflects management and radiation-driven demand
rather than a limiting resource, further reducing its interpretive value
once RH and PAR are included. Finally, CO\textsubscript{2}, RH, and PAR
are centered and scaled within each greenhouse so that a one-unit change
corresponds to a deviation of one standard deviation from the mean of that house.

\subsection{Yield states}

\begin{figure}[H]
\caption{Raw daily fruit weight for the six crop lines in GH-3. Horizontal dashed red lines mark the GH-3 tertile thresholds that define the three yield states—State 1 (low, $<0.55$ kg), State 2 ($0.55$–$1.15$ kg), and State 3 (high, $>1.15$ kg). These cut-points are computed from GH-3 and
used unchanged across greenhouses.}
\label{fig:kgGH3}
  \centering\includegraphics[width=\linewidth]{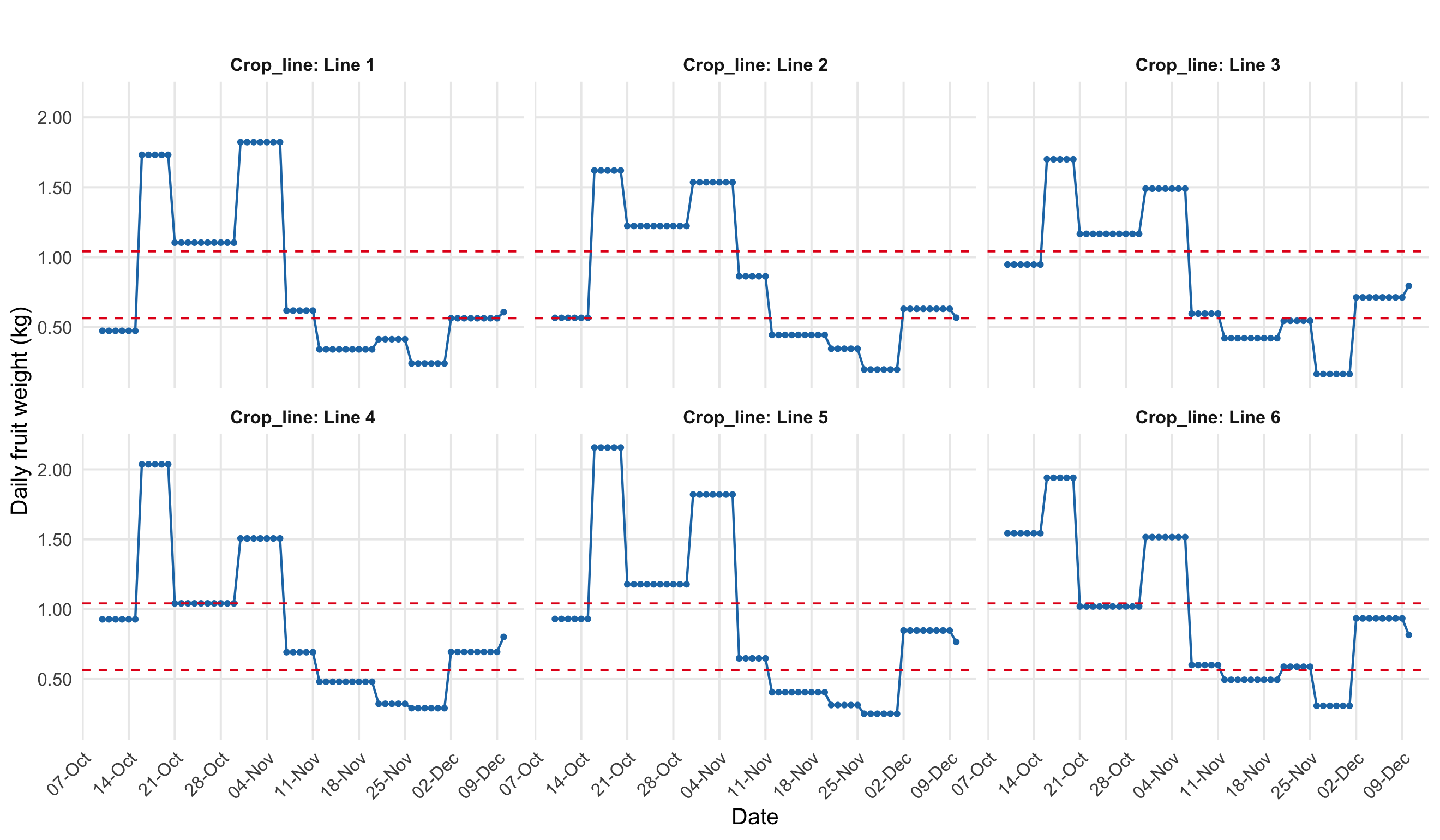}
\end{figure}

The daily fresh weights are clustered around three plateaus (Figure~\ref{fig:kgGH3}). 

Using the GH-3 tertiles ($<0.55$ kg, 0.55–1.15 kg, $>1.15$ kg) we define three levels of production in terms of fresh weight: 

\[
\text{State 1: low},\quad
\text{State 2: medium},\quad
\text{State 3: high}.
\]

The cut-points were computed from GH-3 (control) data and then applied unchanged to all houses so that “low / medium / high” retains the same agronomic meaning across treatments. A three-state partition is wide enough to absorb measurement noise yet captures biologically meaningful jumps (e.g., a line roughly doubling its daily yield).
Applying these thresholds yields nearly even occupancy (small imbalances reflect ties and missing harvest days). Counts (number of line-days) and percentages are reported in table~\ref{tab:statefreq}.

\begin{table}[H]
\centering\small
\caption{Daily state counts by dataset.}
\label{tab:statefreq}
\begin{tabular}{@{}lccc@{}}
\toprule
 & State 1 (low) & State 2 (medium) & State 3 (high) \\
\midrule
GH-3 only (6 lines, 62 d; $n=372$)  & 121 (32.5\%) & 123 (33.1\%) & 128 (34.4\%) \\
All greenhouses pooled ($n=1{,}488$) & 471 (31.7\%) & 486 (32.7\%) & 531 (35.7\%) \\
\bottomrule
\end{tabular}
\end{table}

\begin{figure}[H]
        \centering
        \caption{Each jar of cucumbers represents one of the three daily-yield classes derived from GH-3 tertiles: State 1 (low, $<$0.55 kg), State 2 (medium, 0.55–1.15 kg) and State 3 (high, $>$1.15 kg). Double-headed arrows indicate that transitions are allowed in both directions.}
        \label{fig:flowchart}
        \includegraphics[width=.9\linewidth]{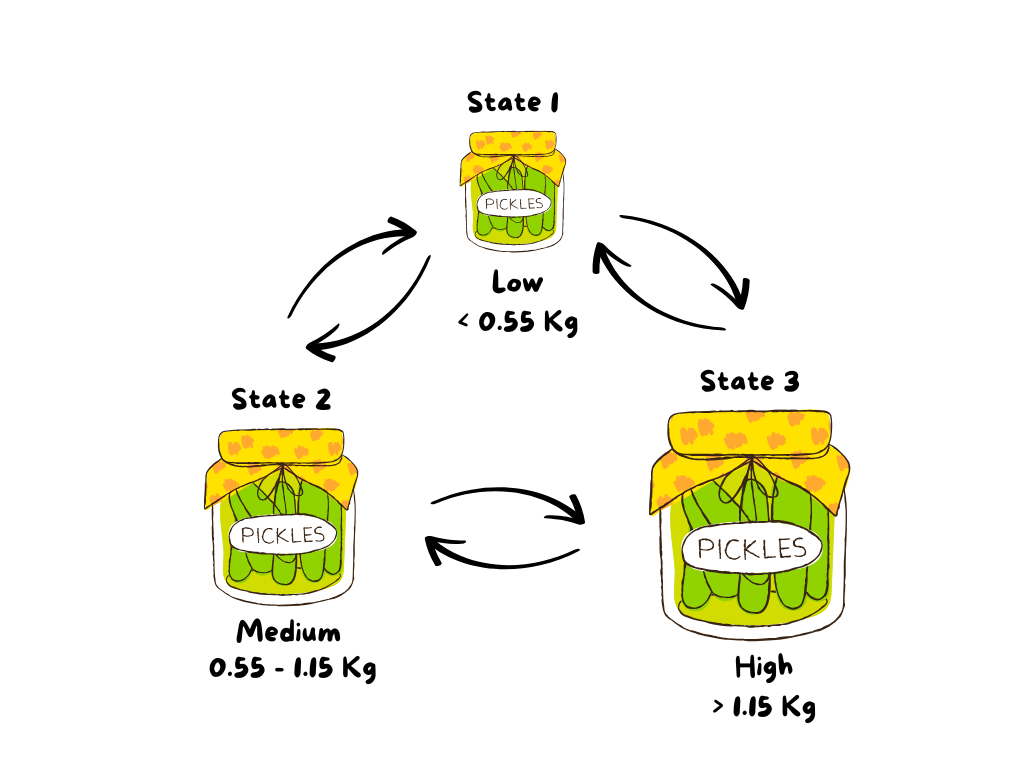}
    \end{figure}

The diagram in Figure~\ref{fig:flowchart} illustrates the conceptual state space with all bidirectional moves. In the empirical analysis, direct transitions $1\leftrightarrow3$ were structurally forbidden (set to zero in $\mathbf Q$) because such jumps were not observed and their inclusion caused numerical instability; reaching State~3 from State~1 occurs via State~2.

These prepared data feed the continuous-time multi-state model
described next.

\section{Methodology}
\label{sec: method}

Multi-state (MS) models are a standard framework for describing the evolution of event-history data in which an experimental unit can occupy one of a finite set of states that may change over time.
These models offer a predictive focus, enabling a forward-looking perspective and informed decision-making based on estimated probabilities. 
A multi-state model is defined as a model for a stochastic process, which at any time occupies one of a set of discrete states. 
In medical studies the states often correspond to disease stages (e.g., healthy $\rightarrow$ ill $\rightarrow$ dead).  
Here, the states represent ordered \emph{daily-yield classes} for each crop line, obtained by discretising the continuous cucumber weight (Section~\ref{sec: Data}). 

The state structure specifies the states and which transitions from state to state are possible. The full statistical model specifies the state structure and the form of the hazard function (intensity function) for each possible transition \citep{hougaard1999multi}.
Multistate models are a well-established method in the medical, biological and engineering fields \citep{Jackson2003}, but they are innovative in greenhouse agronomy.
By quantifying how micro-climate covariates accelerate or delay transitions between low, medium and high yield, growers gain a forward-looking tool to prioritise control actions that most effectively shorten the time to peak production.  Figure~\ref{fig:flowchart} sketches the three-state structure used here; arrows indicate allowed moves, and no state is absorbing, reflecting the possibility of both improvement and decline during the cycle.

Following the notation of \cite{msm}, let $S(t)$ denote the state
occupied by a crop line at (continuous) time $t$.  Conditional on the present state, the process may jump to any other state with \emph{transition
intensity} 
\begin{equation}
  q_{rs}\!\bigl(t,\,\mathbf z(t)\bigr)
  \;=\;
  \lim_{\Delta t\to0}\frac{\Pr\{S(t+\Delta t)=s \mid S(t)=r\}}
                            {\Delta t},
  \qquad r\neq s,
\end{equation}
where $\mathbf z(t)$ collects time-varying, line-specific covariates
(\,centred, scaled $\mathrm{CO_2}$, RH and PAR\,).  
All intensities are assembled in the
$3\times3$ generator matrix $\mathbf Q$, whose rows sum to
zero so that $q_{rr}= -\sum_{s\ne r}q_{rs}$:

\[
\mathbf Q \;=\;
\begin{bmatrix}
-(q_{12}+q_{13}) & q_{12} & q_{13} \\
q_{21} & -(q_{21}+q_{23}) & q_{23} \\
q_{31} & q_{32} & -(q_{31}+q_{32})
\end{bmatrix}.
\]

The off-diagonal entries represent instantaneous risks of moving between the ordered \emph{yield states},
while the diagonals control the exponentially distributed sojourn times.  For example, the expected residence time in state 1 equals $(-q_{11})^{-1}$ and, upon departure, the probability of a direct leap to state 3 is $q_{13}/(-q_{11})$.

Maximum likelihood estimation is performed using the BFGS quasi-Newton algorithm as implemented in the \texttt{msm} R package. For robustness, we also fitted the model with the Nelder–Mead simplex; BFGS consistently achieved a higher log-likelihood and lower AIC for both GH-3 and the pooled model (see Section~\ref{sec: Analysis}). We treat the data as panel observations of an underlying continuous-time Markov chain, using the matrix exponential to link intensities to transition probabilities over the observed day-to-day intervals. Under the Markov assumption, the future evolution depends only on the current state and covariates, not on past history \citep{cox1977theory}. All crop lines are observed throughout the 62-day window, so there is no terminal-event censoring.

Exponentiating the estimated regression coefficients yields hazard ratios (HRs), i.e.\ multiplicative changes in the transition rates per $+1$\,SD increase in a covariate. HRs provide an action-oriented reading: a value of $\text{HR}=2$ for RH on the $1\!\to\!2$ transition, for instance, means that a humid day halves the expected time for a low-yield line to escape the low-yield state.

\section{Analysis and results}
\label{sec: Analysis}

All the analysis have been carried out using the \texttt{msm} R package \citep{msm}. In principle all pairwise transitions are admissible. However, in our data direct jumps between the lowest and highest yield classes were essentially absent. To avoid poorly identified parameters and a near–singular Hessian, we therefore estimated a \emph{constrained} generator that disallows these rare paths,
\[
\mathbf Q_{\text{constr}} \;=\;
\begin{bmatrix}
-\;q_{12} & q_{12} & 0 \\
q_{21} & -\!(q_{21}+q_{23}) & q_{23} \\
0 & q_{32} & -\;q_{32}
\end{bmatrix},
\qquad q_{13}=q_{31}=0,
\]
so that movement from State~1 to State~3 proceeds via State~2. The expected sojourn time in state $r$ remains $(-q_{rr})^{-1}$; under the constraint the probability of an immediate leap $1\!\to\!3$ is zero by construction, though being in State~3 after 30 days is still possible via $1\!\to\!2\!\to\!3$.  

Models were estimated by maximum likelihood using BFGS. For robustness we also fitted the identical specification with the Nelder–Mead simplex. BFGS delivered a systematically better fit (higher log-likelihood, lower AIC) with a well-behaved Hessian:

\begin{table}[H]\centering
\caption{Optimisation sensitivity (constrained $1\leftrightarrow3$): log-likelihood and AIC by algorithm.}
\begin{tabular}{lrr}
\toprule
 & \multicolumn{1}{c}{logLik} & \multicolumn{1}{c}{AIC} \\
\midrule
GH-3, BFGS          & $-63.18$  & $158.37$ \\
GH-3, Nelder--Mead  & $-67.95$  & $167.90$ \\
All GH, BFGS        & $-364.34$ & $784.67$ \\
All GH, Nelder--Mead& $-365.84$ & $787.68$ \\
\bottomrule
\end{tabular}
\end{table}

For the BFGS fits the observed-information Hessian was positive definite (GH-3: $\lambda_{\min}=0.056$, $\lambda_{\max}=106.35$, condition number $\kappa\approx1.89\times10^{3}$; pooled: $\lambda_{\min}=2.07$, $\lambda_{\max}=216.86$, $\kappa\approx1.05\times10^{2}$), indicating satisfactory local identifiability under the constrained specification.

\subsection*{GH-3 (control): single-house results}

\begin{figure}[H]
  \centering
  \caption{GH-3 — Estimated transition intensities $q_{rs}$. Off–diagonals are instantaneous movement rates (day$^{-1}$); diagonals equal $-\sum_{s\neq r}q_{rs}$. Long jumps $1\leftrightarrow3$ are constrained to zero.}
  \label{fig:gh3-intensity}
  \includegraphics[width=.62\linewidth]{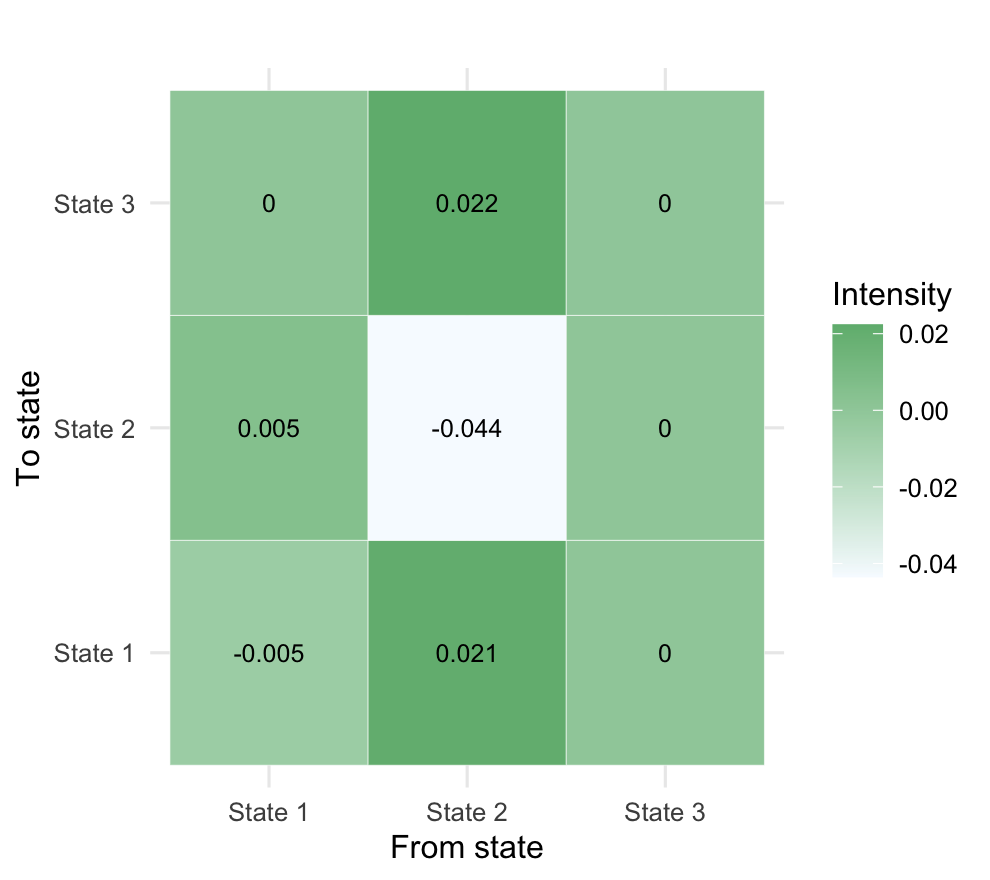}
\end{figure}

Using the constrained generator with $q_{13}=q_{31}=0$, the GH-3 fit shows an ordered, “through–the–middle’’ dynamic. 
The estimated transition intensities (Figure~\ref{fig:gh3-intensity}) imply that state~2 is the main gateway: $q_{21}\approx0.021$ and $q_{23}\approx0.022~\text{day}^{-1}$, so the expected sojourn in state~2 is about $1/(0.021+0.022)\approx23$ days. Escaping state~1 is comparatively slow ($q_{12}\approx0.005~\text{day}^{-1}$, an average stay near $200$ days under baseline conditions), which explains why many low–yield episodes persist unless the microclimate is favourable. The downward intensity from state~3 is extremely small (rounds to $0$ at three decimals), meaning that once a line reaches high yield it very rarely drops within the next month.

\begin{figure}[H]
  \centering
  \caption{GH-3 — Predicted state after 30 days (rows) by current state (columns). Numbers are probabilities.}
  \label{fig:gh3-30d}
  \includegraphics[width=.62\linewidth]{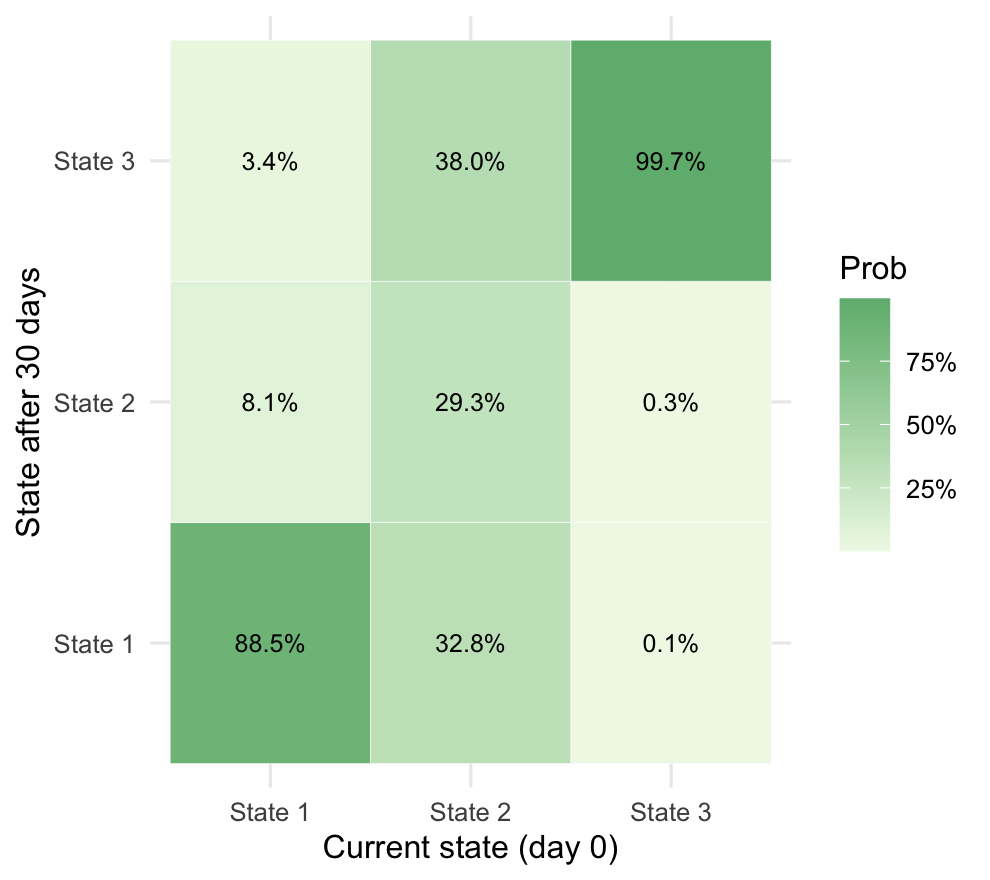}
\end{figure}

Thirty–day transition probabilities translate these rates into forecasts (Figure~\ref{fig:gh3-30d}). If a line is low today (state~1), the model predicts it will still be low after 30 days with probability $88.5\%$, move to medium with $8.1\%$, and to high with $3.4\%$. Starting from medium (state~2), trajectories are more fluid: $32.8\%$ fall to low, $29.3\%$ remain medium, and $38.0\%$ climb to high. Once in the high state, persistence dominates: the probability of remaining high at 30 days is $99.7\%$.

\begin{figure}[H]
  \centering
  \caption{GH-3 — Hazard ratios (log scale on $y$) for the two forward transitions per $+1$\,SD in each covariate. Bars above/below $1$ indicate acceleration/deceleration.}
  \label{fig:gh3-hr}
  \includegraphics[width=.70\linewidth]{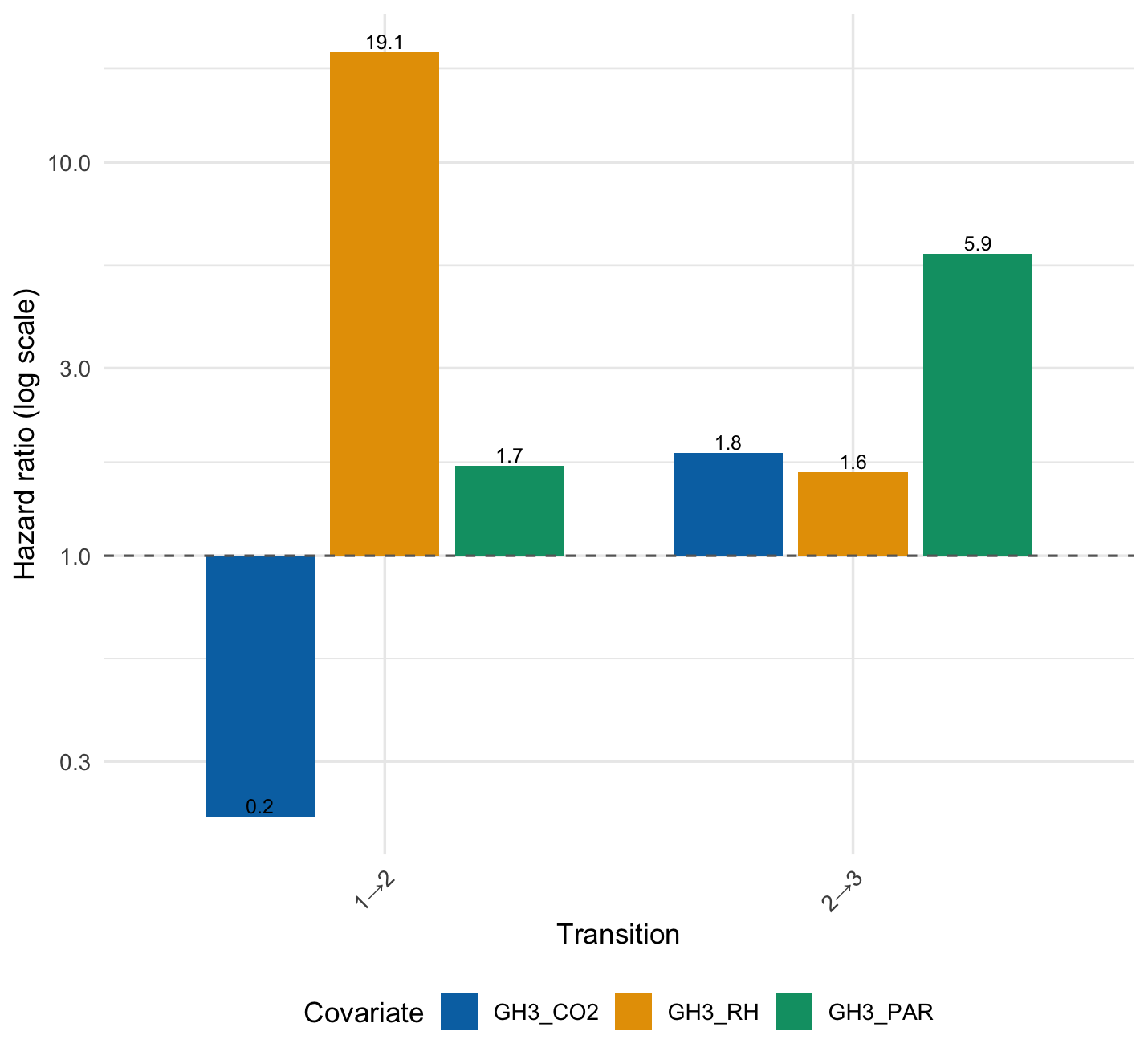}
\end{figure}

Microclimate effects are expressed as hazard ratios (HR) per $+1$\,SD change (Table~\ref{tab:gh3-HRtable} and Figure~\ref{fig:gh3-hr}). 

\begin{table}[H]
\centering\small
\caption{GH--3 (control). Hazard ratios (HR) per $+1$\,SD change in covariates from the constrained three-state model ($q_{13}=q_{31}=0$). 
\textbf{Bold} entries have 95\% CIs excluding 1. 
Cells marked with~$\dagger$ correspond to sparse transitions (very few observed events), for which estimates may sit at the boundary and CIs can be unstable or degenerate; interpret with caution.}
\label{tab:gh3-HRtable}
\begin{tabular}{@{}lccc@{}}
\toprule
Transition & CO$_2$ & RH & PAR \\
\midrule
$1\to2$ 
  & \textbf{0.22}\;(0.07--0.72) 
  & \textbf{19.06}\;(1.03--351.24) 
  & 1.69\;(0.76--3.79) \\
$2\to1$ 
  & 2.66\;(0.98--7.21) 
  & 0.43\;(0.07--2.85)   
  & 0.40\;(0.02--8.43) \\
$2\to3^\dagger$ 
  & 1.83\;(0.30--11.01) 
  & 1.63\;(0.34--7.78)  
  & 5.86\;(0.94--36.52) \\
$3\to2^\dagger$ 
  & 0.00\;(0.00--2.15)  
  & 0.00\;(0.00--0.00)  
  & 0.00\;(0.00--0.05) \\
\bottomrule
\end{tabular}
\end{table}

Wide intervals reflect the small number of observed transitions for certain moves (particularly $2\to3$ and $3\to2$).
Relative humidity is the main lever for escaping the low state: $\text{HR}_{1\to2}=19.1$ with a wide 95\% CI ($\approx1.0$–$351$), indicating acceleration out of state~1 when conditions are humid. The breadth of this interval reflects limited information for that transition: few $1\to2$ events occur and the likelihood is relatively flat, so sampling variability is large. In such cases we focus on direction (HR$>$1 indicates acceleration) rather than the exact multiplier. PAR supports upward movement at both steps, moderately from $1\to2$ (HR $=1.69$, 95\% CI $\approx0.76$–$3.79$) and strongly from $2\to3$ (HR $=5.86$, 95\% CI $\approx0.94$–$36.5$). The second interval is again wide because transitions into state~3 are comparatively rare; nonetheless the point estimate and the 30-day probabilities consistently identify light as the key driver of entering peak production. CO$_2$ shows a nuanced pattern: higher CO$_2$ days are associated with a slower $1\to2$ move (HR $=0.22$, 95\% CI $\approx0.07$–$0.72$) and a weak, imprecise acceleration for $2\to3$ (HR $=1.83$, 95\% CI $\approx0.30$–$11.0$). Within GH-3 this likely captures management correlations (e.g., dosing and ventilation coinciding with light and humidity) rather than a pure physiological effect, and we refrain from strong quantitative claims when intervals straddle the null.

Practically, the GH-3 model suggests: (i) low-yield spells are “sticky’’ under baseline conditions but can be shortened by raising RH (lowering VPD) and, to a lesser extent, by increasing PAR; (ii) progression from the mid-yield plateau to the high state is primarily light-limited; and (iii) once in the high state, lines tend to remain there over a month. Where confidence intervals are very wide because events are scarce, we report point estimates and CIs for completeness but base interpretation on the sign and consistency with the transition probabilities; the pooled analysis below partly mitigates this imprecision by borrowing information across houses.

\subsection*{All greenhouses pooled}

Pooling the four houses serves two goals: it increases information (24 lines, $1{,}488$ line–days) to sharpen inference on microclimate effects, and it tests whether a single dynamical scheme can describe all houses once baseline differences are absorbed by fixed effects (GH-3 as reference). Covariates are $z$–scores computed \emph{within} house so their slopes are directly comparable. As motivated above, we retain the constrained generator with $q_{13}=q_{31}=0$, reflecting the empirical absence of direct low–high jumps.

Fitting with BFGS yielded a higher likelihood and lower AIC than Nelder–Mead (logLik $=-364.34$ vs.\ $-365.84$; AIC $=784.67$ vs.\ $787.68$), and the Hessian was well conditioned (smallest eigenvalue $\approx2.07$), indicating a stable optimum.

\begin{figure}[H]
  \centering
  \caption{Pooled-model baseline transition intensities $q_{rs}$ (day$^{-1}$) at centred covariates.}
  \label{fig:pool-intensity}
  \includegraphics[width=.58\linewidth]{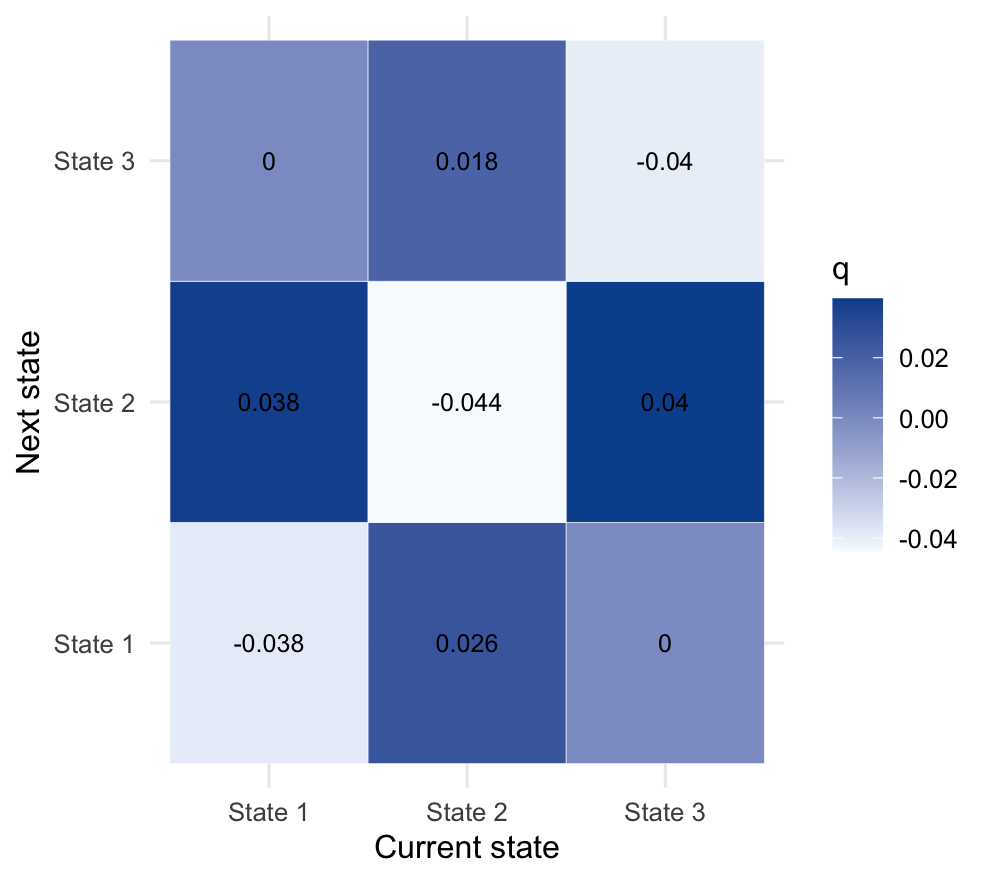}
\end{figure}

At centred covariates the estimated intensities show an overall upward drift (Figure~\ref{fig:pool-intensity}): $q_{12}\!\approx\!0.038$ and $q_{23}\!\approx\!0.040$ day$^{-1}$ exceed the corresponding downward rates $q_{21}\!\approx\!0.026$ and $q_{32}\!\approx\!0.018$. These imply mean sojourns of roughly $26$ days in state~1, $23$ in state~2 and $25$ in state~3.

\begin{figure}[H]
  \centering
  \caption{Thirty-day transition probabilities for the pooled model (columns: state today; rows: state in 30 days).}
  \label{fig:pool-30d}
  \includegraphics[width=.58\linewidth]{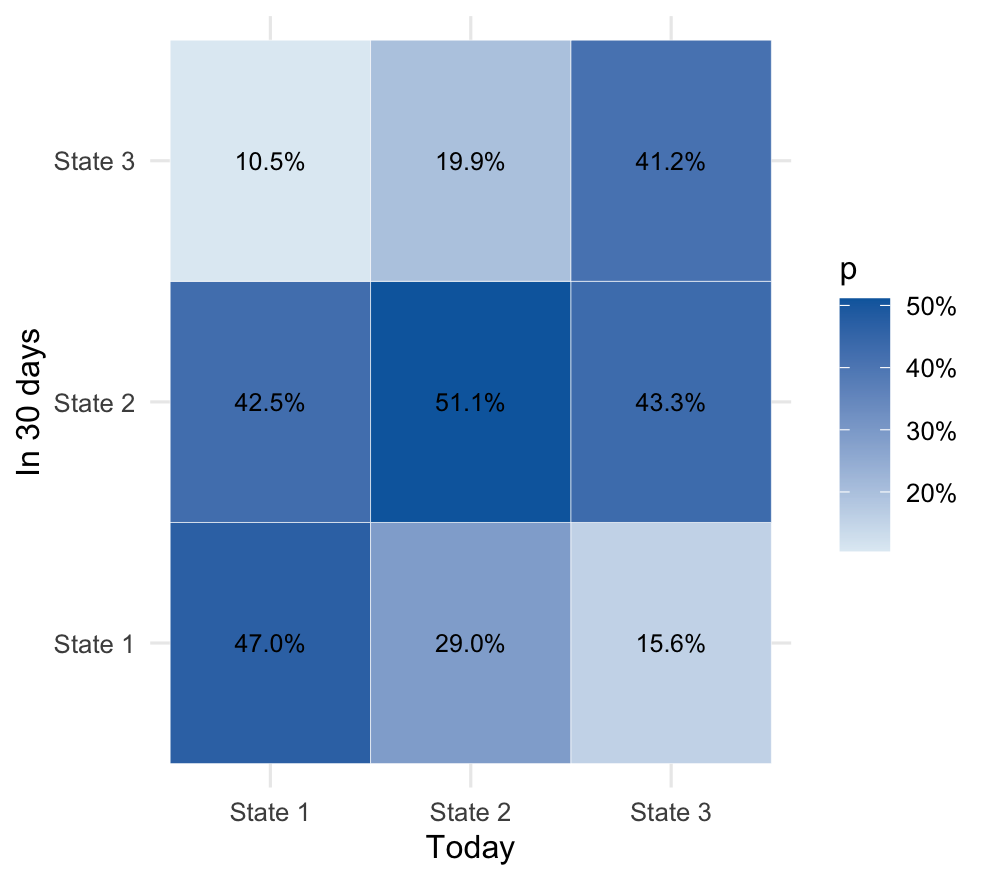}
\end{figure}

Translating intensities into a one-month outlook (Figure~\ref{fig:pool-30d}), a plot in state~1 today has a $47.0\%$ chance to remain low after 30 days, $42.5\%$ to move to state~2 and $10.5\%$ to reach state~3. From state~2 the most likely outcome is to stay in~2 ($51.1\%$), with $29.0\%$ dropping to~1 and $19.9\%$ rising to~3. From state~3 the peak is comparatively short-lived: only $41.2\%$ remain in~3 after a month, while $43.3\%$ are expected to be in~2 and $15.6\%$ in~1 via the $3\!\to\!2$ (and possibly $2\!\to\!1$) path.

\begin{figure}[H]
  \centering
  \caption{Pooled-model hazard ratios (HR) for forward transitions $1\!\to\!2$ and $2\!\to\!3$. 
  Bars show multiplicative changes in transition rates (values $>$1 speed the move; $<$1 slow it). 
  PAR, RH and CO\textsubscript{2} are within-house $z$-scores, so HRs are per $+1$\,SD day relative to that house’s mean. 
  GH-2, GH-4 and GH-5 are house indicators and represent baseline multipliers \emph{relative to GH-3} after adjusting for microclimate; they should be interpreted as residual structural/management contrasts rather than direct microclimate effects.}
  \label{fig:pool-hr}
  \includegraphics[width=.70\linewidth]{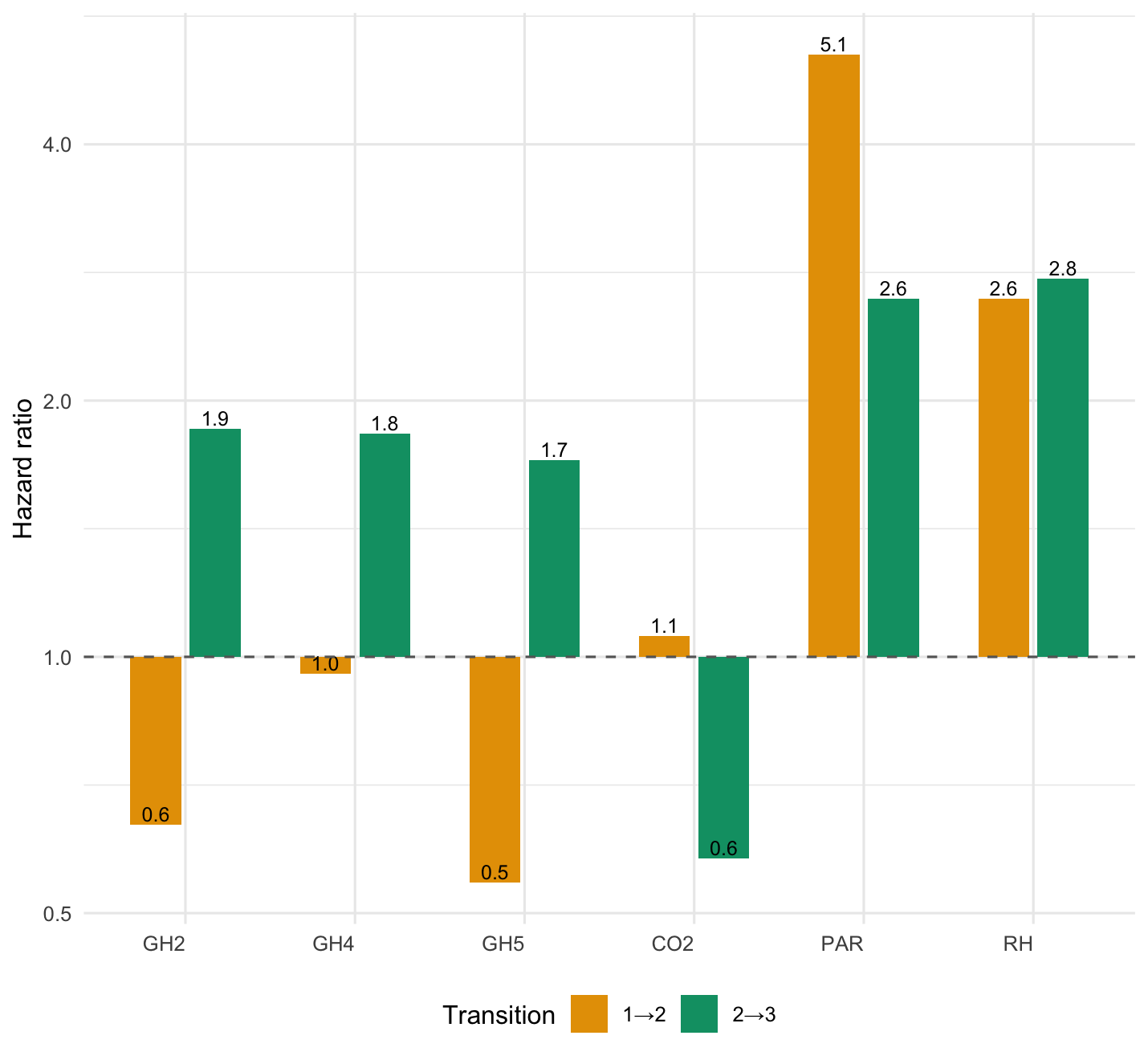}
\end{figure}

Microclimate effects (Tables~\ref{tab:pooled-HR-microclimate} and~\ref{tab:pooled-HR-GH}, and Figure~\ref{fig:pool-hr}) are coherent and agronomically large. 

\begin{table}[H]
\centering\small
\caption{Pooled model (all greenhouses). Hazard ratios (HR) per $+1$\,SD change in microclimate covariates from the constrained three-state model ($q_{13}=q_{31}=0$). \textbf{Bold} entries have 95\% CIs excluding 1.}
\label{tab:pooled-HR-microclimate}
\begin{tabular}{@{}lccc@{}}
\toprule
Transition & CO$_2$ & RH & PAR \\
\midrule
$1\to2$ & 1.06\;(0.72--1.55) & \textbf{5.10}\;(2.55--10.21) & \textbf{2.63}\;(1.71--4.05) \\
$2\to1$ & 0.82\;(0.57--1.18) & \textbf{0.41}\;(0.22--0.77)  & \textbf{0.12}\;(0.04--0.31) \\
$2\to3$ & 0.58\;(0.25--1.36) & \textbf{2.64}\;(1.28--5.42)  & \textbf{2.78}\;(1.72--4.50) \\
$3\to2$ & 1.35\;(0.48--3.76) & \textbf{0.13}\;(0.05--0.33)  & \textbf{0.27}\;(0.13--0.56) \\
\bottomrule
\end{tabular}
\end{table}

\begin{table}[H]
\centering\small
\caption{Pooled model (all greenhouses). Greenhouse fixed effects (HR) relative to GH--3, adjusted for CO$_2$, RH, and PAR, from the constrained three-state model ($q_{13}=q_{31}=0$). All 95\% CIs include 1, indicating no clear baseline difference in state-transition rates across houses once microclimate is accounted for.}
\label{tab:pooled-HR-GH}
\begin{tabular}{@{}lccc@{}}
\toprule
Transition & GH--2 vs GH--3 & GH--4 vs GH--3 & GH--5 vs GH--3 \\
\midrule
$1\to2$ & 0.64\;(0.21--1.91) & 0.96\;(0.36--2.57) & 0.54\;(0.19--1.53) \\
$2\to1$ & 1.12\;(0.39--3.22) & 0.97\;(0.34--2.81) & 0.81\;(0.27--2.42) \\
$2\to3$ & 1.85\;(0.50--6.88) & 1.83\;(0.60--5.58) & 1.70\;(0.45--6.46) \\
$3\to2$ & 1.68\;(0.49--5.77) & 1.44\;(0.54--3.82) & 0.72\;(0.19--2.76) \\
\bottomrule
\end{tabular}
\end{table}

A $+1$\,SD day in \textbf{RH} substantially \emph{accelerates} upward moves and \emph{dampens} regressions: $\mathrm{HR}_{1\to2}=5.10$ (95\% CI 2.55–10.21) and $\mathrm{HR}_{2\to3}=2.64$ (1.28–5.42), while the reverse hazards shrink to $\mathrm{HR}_{2\to1}=0.41$ (0.22–0.77) and $\mathrm{HR}_{3\to2}=0.13$ (0.05–0.33). \textbf{PAR} shows the same qualitative pattern, $\mathrm{HR}_{1\to2}=2.63$ (1.71–4.05) and $\mathrm{HR}_{2\to3}=2.78$ (1.72–4.50), with downward moves slowed to 0.12 (0.04–0.31) and 0.27 (0.13–0.56). Interpreted as waiting times, these HRs imply that brighter, more humid days reduce the expected time to leave a state by roughly 60–75\%. By contrast, day-to-day \textbf{CO\textsubscript{2}} deviations show no clear pooled signal ($\mathrm{HR}_{1\to2}=1.06$, 0.72–1.55; $\mathrm{HR}_{2\to3}=0.58$, 0.25–1.36).

The \textbf{house fixed effects} compare each treatment to GH-3 after controlling for within-house–standardised PAR, RH and CO\textsubscript{2}, so they capture residual baseline differences (e.g., structural or management features) rather than mean microclimate. 
Point estimates suggest GH-2 and GH-5 are somewhat slower to exit low yield than GH-3 ($\mathrm{HR}_{1\to2}=0.64$ and $0.54$), while all three houses show slightly faster progression from medium to high yield (HRs 1.70–1.85). However, confidence intervals are wide and include 1, so the data remain compatible with no material difference in baseline dynamics across treatments; the dominant drivers of movement between states are PAR and RH.

\section{Discussion}
\label{sec: Discussion}

The three–state MSM turns noisy sensor streams into time–to–event quantities that a grower can act on: expected sojourn in each yield class and short–horizon probabilities of upgrading or downgrading given today’s microclimate. Because covariates are centred and scaled within house, hazard ratios read as elasticities to day–to–day departures from the usual climate, which makes them directly usable in control logic.

Two operational levers emerge. \textit{Humidity management} (high RH / low VPD) both accelerates upgrades and cushions against regressions; it can therefore serve as a stabiliser under adverse spells (e.g., dim, dry weeks) by temporarily lifting RH set–points, relaxing ventilation windows, or deploying fogging to suppress $3\!\to\!2$ and $2\!\to\!1$ hazards. 
\textit{Light availability} is the primary driver for entering and sustaining the high–yield state; when supplemental lighting is available, allocating hours to lines closest to the $2\!\to\!3$ threshold should yield the largest reduction in time–to–peak. In practice, the MSM can rank lines by expected days to upgrade under current RH and PAR and guide scarce resources to the blocks with the steepest expected gains.
These patterns are consistent with greenhouse cucumber studies in which total solar radiation is the dominant driver of yield, and with the classic rule-of-thumb that a 1\% increase in light yields $\approx1\%$ more product \citep{Ding_2019_hortsci, marcelis2005quantification, marcelis1998modelling}.

Pooling houses indicates that, once RH and PAR are accounted for, baseline differences across treatments are modest over the short horizon considered. This shifts emphasis from structural labels (e.g., PV roof or \mbox{CO$_2$} regime) to daily steering: nights and mornings with higher RH and days with higher PAR are the conditions that reliably shorten time to upgrade and buffer against downgrades across all houses. 
Structural interventions still matter insofar as they reshape the distribution of RH and PAR days; the MSM quantifies their effective value by how much they shift the transition probabilities experienced over a season.
This emphasis on radiation, temperature and humidity as primary levers aligns with overviews of closed/semi-closed greenhouses, where climate control strategies are explicitly tied to crop performance \citep{de2012overview}.

Uncertainty warrants explicit handling. Some transitions are rare within a single house, which yields wide confidence intervals for the associated hazard ratios. In such cases, decisions should lean on the pooled estimates, which borrow strength across houses, and on the robust qualitative pattern (RH and PAR accelerate upgrades and decelerate regressions) rather than on precise multipliers. Constraining empirically absent $1\!\leftrightarrow\!3$ jumps improved identifiability without limiting practical forecasts, since $1\!\to\!3$ remains attainable via $1\!\to\!2\!\to\!3$.


\section{Conclusions and future developments}
\label{sec: conclusions}

This study reinterprets greenhouse cucumber production as movement across three ordered yield states and estimates those dynamics with a continuous–time multistate model (MSM) whose transition rates depend on within-greenhouse $z$–scores of RH, PAR and CO\textsubscript{2}, plus greenhouse fixed effects. 
The trial was carried out in four contiguous greenhouses located in Volos, Greece, from 10~Oct to 10 Dec 2024, with data collected daily. 
All greenhouses share the same geometry, hydroponic gutters, and climate controller, but differ one factor under study from the “control” greenhouse GH-3.
Defining the states once from the GH–3 tertiles ensures agronomic comparability across houses and anchors the interpretation of ``low'', ``medium'' and ``high'' productivity. Across both single–house and pooled fits, a coherent pattern emerges: there is an inherent upward drift through the medium state; low–yield spells tend to persist unless conditions improve; and high–yield episodes are comparatively stable over short horizons.

Beyond describing patterns, the MSM provides quantities that growers can act on. Transition intensities and their covariate effects translate sensor streams into expected sojourn times and short–horizon upgrade/downgrade probabilities. Because covariates are centred and scaled within house, hazard ratios read as elasticities to feasible, day–to–day departures from the usual climate. In practice, this enables simple rules of thumb that are both defensible and implementable: \emph{raise RH to accelerate exits from low yield and to cushion against regressions}, and \emph{supplemental light to lines closest to the $2\!\to\!3$ threshold, shorten time–to–peak}. 

When the four houses are pooled, baseline differences across treatments (CO\textsubscript{2} enrichment, PV roof, and their combination) are modest and statistically uncertain once RH and PAR are accounted for over the 62–day window. This shifts emphasis from structural labels to daily steering: days with higher RH and days with higher PAR are the conditions that reliably shorten time to upgrade and buffer against downgrades across all houses. 
Structural interventions still matter to the extent that they reshape the distribution of RH and PAR days; the MSM provides a natural way to value those interventions by simulating how they shift transition probabilities experienced over a season.

Several limitations qualify these conclusions. Weekly harvests were forward–carried to daily values, which smooths abrupt changes and may understate very fast transitions. The 62–day window motivates a time–homogeneous, Markov specification; if duration dependence or seasonal drift is important, semi–Markov or piecewise–constant intensities would be preferable. Direct $1\!\leftrightarrow\!3$ jumps were constrained absent in the data; this improved identifiability without limiting practical forecasting because $1\!\to\!3$ remains attainable via $1\!\to\!2\!\to\!3$. Finally, microclimate variables can correlate with management actions (e.g., venting, dosing, screening), so effects—particularly for CO\textsubscript{2}—should be interpreted within that context.
The CO\textsubscript{2} result merits nuance. Although our day–to–day CO\textsubscript{2} deviations showed no clear effect once RH and PAR were included, targeted model–based enrichment can increase cucumber yield substantially ($\approx 35\%$ reported by \cite{klaring2007model}), and semi–closed houses may exhibit vertical temperature and humidity gradients that reshape canopy microclimate. Factors that motivate richer CO\textsubscript{2} control data and vertical profiles in future studies \citep{qian2010vertical}.

These caveats naturally define a roadmap. Methodologically, longer horizons and richer structure; time–varying or semi–Markov intensities to capture duration/seasonality, hierarchical random effects to stabilize sparse transitions, and interaction or nonlinearity (e.g., RH$\times$PAR, splines, or threshold effects), would improve both fit and interpretability. From an operational standpoint, tighter coupling to short–term weather forecasts and online updating would enable rolling scenario analysis; embedding the MSM within a decision layer (e.g., model–predictive control with explicit costs/constraints) would turn probabilities into set–point recommendations with quantified trade–offs. Finally, integrating the MSM layer with existing microclimate simulators and control policies would deliver a lightweight yet interpretable building block for greenhouse digital twins: one that ranks lines by upgrade risk, quantifies the expected return of feasible set–point changes, and communicates uncertainty alongside recommendations.


\section*{CRediT authorship contribution statement}
\textbf{Emiliano Seri:} Writing – original draft \& editing, Visualization, Validation, Software, Methodology, Investigation, Formal analysis, Data curation, Conceptualization. 
\textbf{Francesco Biso:} Writing – original draft \& editing, Conceptualization, Investigation, Formal analysis, Methodology, Data curation. \textbf{Gianluigi Bovesecchi:}  Writing – review \& editing, Data curation. 
\textbf{Nikolaos Katsoulas:} Writing – review \& editing, Data curation. 
\textbf{Cristina Cornaro:} Writing – review \& editing, Conceptualization, Project administration, Supervision.


\section*{Data availability}

The authors do not have permission to share data.

\section*{Declaration of competing interest}
The authors declare that they have no known competing financial interests or personal relationships that could have appeared to influence the work reported in this paper.


\section*{Acknowledgments}

This research was carried out in the framework of the REGACE project funded by the European Union under Grant Agreement No 101096056. Views and opinions expressed are however those of the author(s) only and do not necessarily reflect those of the European Union or CINEA. Neither the European Union nor the granting authority can be held responsible for them.

The crop and greenhouse microclimate data used in this work have been collected in the frame of the Horizon project REGACE. The authors would like to thank Mrs Sofia Faliagka, Mr Ioannis Naounoulis, and Mrs Evangelia Skantzou for their participation in the REGACE project and Prof. Chrysoula Papaioannou for the coordination of the project for the University of Thessaly.


\bibliographystyle{model1-num-names}
\bibliography{MAIN}


\end{document}